# CODYRUN, OUTIL DE SIMULATION ET D'AIDE A LA CONCEPTION THERMO-AERAULIQUE DE BATIMENTS

**Harry BOYER , Alain BASTIDE, Philippe LAURET**

LGI – Equipe GCTH – Université de La Réunion  – IUT de Saint Pierre, Département Génie Civil, 40 Avenue de Soweto, 97410 Saint Pierre – Ile de la Réunion

*RESUME. Cet article présente le code de simulation CODYRUN développé à l'Université de La Réunion. Il s'agit d'un logiciel de simulation thermique multizone, intégrant un module aéraulique et un modèle hydrique. Une de ses particularités est celle d'être simultanément un outil de recherche autour duquel beaucoup de nos travaux se sont articulés (plateforme d'intégration de modèles, validation, ...), un outil d'aide à la conception utilisé par le laboratoire et des professionnels et enfin un outil pédagogique utilisé en enseignement à plusieurs niveaux. Après une présentation du caractère multimodèles de l'application dans une première section, les trois modules liés à la physique des phénomènes intégrés sont explicités. Des éléments de la validation sont évoqués dans la section suivante et, enfin, quelques éléments de l'interface du code sont donnés.*

*MOTS-CLÉS : CODYRUN, simulation, modélisation.*

*ABSTRACT. This article presents the CODYRUN software developped by University of La Réunion. It is  a multizone thermal software, with detailled airflow and humidity transfer calculations. One of its specific aspects is that it constitutes a research tool, a design tool used by the lab and professionnals and also a teaching tool. After a presentation of the multiple model aspect, some details of the tree modules associated to physical phenomenons are given. Elements of validation are exposed in next paraghaph, and then a few details of the front end.*

## 1. INTRODUCTION

Cet article présente le code de simulation CODYRUN développé à l'Université de La Réunion et initié, il y a plus de 10 ans, dans le cadre d'une collaboration Université de La Réunion / INSA de Lyon. Il s'agit d'un logiciel de simulation thermique multizones, intégrant un module aéraulique et un modèle hydrique. Au cours du temps, ce logiciel a fortement évolué suivant les nombreuses sollicitations internes et externes au laboratoire. Une de ses particularités est celle d'être simultanément un outil de recherche autour duquel beaucoup des travaux de l'équipe se sont se sont articulés (plateforme d'intégration de modèles, validation, édiction de règles expertes, ...), un outil d'aide à la conception utilisé tant par le laboratoire (dans le cadre de contrats) que par des professionnels (BET, architectes) et enfin un outil pédagogique utilisé en enseignement à plusieurs niveaux d'étude. Après une présentation du caractère multimodèles de l'application, les trois modules liés à la physique des phénomènes intégrés sont détaillés. Des éléments de la validation sont évoqués dans la section suivante suivant et enfin, quelques éléments de l'interface du code sont donnés.

## 2. L'ASPECT MULTI-MODELES DE L'APPLICATION

Cette caractéristique de l'application (Boyer & al., 1998) a fortement contraint l'architecture de l'application en terme de décomposition par blocs, de choix de schémas de résolution et est donc présentée en préalable.

Dès les premières années de diffusion significative des outils de simulation et de conception, l'un des problèmes posés a été celui de leur adéquation a besoins des acteurs du processus de conception. En considérant la façon dont ils sont produits, les modèles thermiques sont le plus souvent obtenus par assemblage de modèles élémentaires, décrivant chacun le comportement thermique des éléments constituant le bâtiment. Une fois les modèles élémentaires choisis (en terme de précision, de temps calcul, de sorties disponibles, de la faisabilité du couplage avec d'autres modèles, ...), cette approche conduit à un modèle du bâtiment figé ou *monolithique*. Il répond précisément au besoin d'un acteur (dimensionnement de systèmes de climatisation par exemple). Ainsi, au vu des besoins très différents des acteurs du domaine (allant du concepteur au physicien du bâtiment), un outil dont les modèles élémentaires sont figés ne saurait répondre à lui seul à l'ensemble des besoins des acteurs du processus de conception. Par exemple, deux objectifs différents sont d'une part d'évaluer une consommation énergétique annuelle et d'autre part d'étudier la réponse (horaire ou sub-horaire) d'un composant du bâtiment sur une séquence climatique donnée. Intuitivement, l'emploi d'un modèle thermique détaillé dans le premier cas conduira à des temps de calculs importants. De même, l'utilisation d'un modèle trop simplifié ne pourra mener qu'à des estimations peu précises, mais rapides, dans le cas de la réponse. Un lien fort existe donc entre notion de qualité de modèle, objectif de la simulation et temps de calcul.

Notre démarche, et de ce fait notre contribution, a été différente. Elle vise à l'intégration d'une bibliothèque de modèles élémentaires interchangeables. Dans la plupart des logiciels de simulation existants, le choix des modèles élémentaires est effectué lors de phase d'analyse et l'usage de ces modèles est globale à l'ensemble du bâtiment (par exemple, le même modèle conductif est enclenché pour toutes les parois opaques). En conséquence, si ces modèles sont un tant soit peu détaillés, les temps de calcul ne sont plus compatibles avec un outil de conception. Pour contourner le compromis entre précision et temps de calcul, il nous est apparu intéressant de permettre, pour certains des phénomènes (ou systèmes), une application sélective des modèles. Il s'agit d'offrir la possibilité de choisir des modèles différents pour des entités de même niveau hiérarchique (les zones ou les parois par exemple). Le but poursuivi est alors d'asservir une entité à un niveau de complexité souhaité. Il est clair que cette démarche sélective doit s'assurer d'un préalable concernant l'intensité des couplages des systèmes de même niveau (et des zones en particulier), préalable qui est du ressort de l'utilisateur expert.

La description d'un bâtiment se fait au travers d'éléments classiques. Ils se décomposent de la manière suivante : le bâtiment, les zones, les séparations entre zones (nommées inter ambiances) et enfin les composants. L'architecture sera détaillée plus en avant. Compte tenu du caractère sélectif des modèles, il est nécessaire de distribuer les informations liées aux modèles dans les structures de description des entités et donc au niveau des fenêtres de l'application (cf par exemple celle du composant paroi, qui fait apparaître des informations liées au modèle conductif)

En terme d'objectif des simulations, la logique précédente conduit à autoriser différents types de simulations multizones, répondant au problème thermique sans aéraulique (bâtiment fermé ou équipé de VMC), thermique avec aéraulique à débit détaillé (modèle en pression), aéraulique ou encore thermique, aéraulique et hydrique. C'est alors une architecture par blocs correspondant à chacun des phénomènes (thermique, aéraulique, hydrique) qui se prête le plus facilement à cette démarche. Les couplages entre ces phénomènes sont gérés de manière itérative, en laissant de surcroît à l'utilisateur expert le choix des options de couplage (par exemple entre les modules thermique et aéraulique) et des critères associés.

## 3. PRESENTATION DES MODULES LIES A LA MODELISATION PHYSIQUE DES PHENOMENES :

### 3.1. PRESENTATION DU MODULE THERMIQUE

Cette partie est détaillée au niveau de la référence (Boyer, 96). Avec les hypothèses classiques d'isothermie du volume d'air des zones, de transferts conductifs unidirectionnels, d'échanges superficiels linéarisés, de l'analyse nodale intégrant les différents modes de transfert (conduction, convection et rayonnement) permettant d'atteindre pour chaque zone thermique constitutive du bâtiment l'ensemble des réponses de l'ambiance (et éléments associés).

Le modèle physique d'une zone est alors obtenu en assemblant les modèles thermiques de chacun des éléments parois, vitrages, volume d'air, qui constituent la zone. Pour fixer les idées, les équations rencontrées du type :

$$C_{si}\frac{dT_{si}}{dt} = h_{ci}(T_{ai} - T_{si}) + h_{ri}(T_{rm} - T_{si}) + K(T_{se} - T_{si}) + \varphi_{swi} \quad (1)$$

$$C_{se}\frac{dT_{se}}{dt} = h_{ce}(T_{ae} - T_{se}) + h_{re}(T_{sky} - T_{se}) + K(T_{si} - T_{se}) + \varphi_{swe} \quad (2)$$

$$C_{ai}\frac{dT_{ai}}{dt} = \sum_{j=1}^{N_w} h_{ci} S_j (T_{ai} - T_{si(j)}) + c\dot{Q}(T_{ae} - T_{ai}) \quad (3)$$

$$0 = \sum_{j=1}^{N_w} h_{ri} A_j (T_{si}(j) - T_{rm}) \quad (4)$$

Les équations de type *(1)* et *(2)* traduisent les bilans thermiques respectifs des nœuds de surface intérieurs et extérieurs. $N_W$ désignant le nombre de parois de l'enveloppe, l'équation *(3)* est celle du bilan thermoconvectif de l'air, compte tenu d'un débit $\dot{Q}$ entre l'intérieur et l'extérieur. *(4)* est l'équation d'équilibre radiatif du nœud de température radiante moyenne.

Le caractère générique de l'application (pas dédiée à un nombre de zones ou à un type de zone particulier) nécessite de porter une attention particulière à la génération du maillage du bâtiment. Dans la même logique que précédemment (application sélective des modèles), vis à vis du multizonage, c'est un procédé de couplage itératif entre les zones qui est implanté. De nouveaux modèles ont été implémenté au cours du temps (et ce depuis la version initiale de l'outil). Ils sont liés, entre autre, à la méthode des radiosités, à la gestion du couplage radiatif (Courte Longueur d'Onde) des zones (à travers les vitrages) ou encore à l'intégration de la réduction du diffus par les masques proches (Lauret, 2001). Pour une étude spécifique liée à l'intégration d'algorithmes de réduction modale (Berthomieu & al., 2003), la mise sous forme canonique d'état (au sens de l'automatique) du système thermique obtenu par zone a été utilisée.

En terme d'utilisation effective, les principales sorties de ce module de thermique sont celles liées aux températures des éléments discrétisés, aux flux surfaciques, aux indices de confort des différentes zones, à la consommation des appareils de climatisation, ...

### 3.2. PRESENTATION DU MODULE AERAULIQUE

Un modèle en pression pour le calcul effectif des débits prend en compte les effets moteurs vent et le tirage thermique. A ce jour, les composants aérauliques intégrés sont les bouches de ventilation, les petites ouvertures (régies par l'équation de crack flow, $\dot{m} = K(\Delta P)^n$) et les grandes ouvertures verticales intérieures. Le modèle (Roldan, 1984) des grandes ouvertures extérieures nous ayant récemment montré ses limites (par comparaison avec la CFD), il sera réputé non intégré, bien qu'il fasse l'objet de développements actuels.

Le bilan massique de chaque zone (en présence de ventilation mécanique) conduit à l'établissement d'un système non linéaire admettant comme inconnues les pressions de référence des zones.

$$\begin{cases} \sum_{i=0, i \neq 1}^{i=N} \dot{m}(i,1) + \dot{m}_{vmc}(1) = 0 \\ \sum_{i=0, i \neq 2}^{i=N} \dot{m}(i,2) + \dot{m}_{vmc}(2) = 0 \\ \quad \ldots \\ \sum_{i=0}^{i=N-1} \dot{m}(i,N) + \dot{m}_{vmc}(N) = 0 \end{cases}$$

$\dot{m}(i,j)$ est le débit (kg/s) de la zone $i$ vers $j$, $\dot{m}_{vmc}(k)$ celui extrait de la zone $k$ et $N$ le nombre total de zones. Après l'établissement automatique de ce système non linéaire, sa résolution est effectuée à l'aide d'une variante de la méthode de Newton Raphson (sous relaxée), mettant en oeuvre l'algorithme de Picard (pour initialiser les valeurs de pression entre 2 itérations) pour promouvoir la convergence.

### 3.3. PRESENTATION DU MODULE HYDRIQUE

En raison du découplage opéré, les températures sèches de chacune des zones ainsi que les débits inter zones ont été préalablement calculés. A ce moment précis du calcul, pour un bâtiment donné, une équation matricielle du type $C_h \dot{r} = A_h r + B_h$ régit l'évolution des humidités spécifiques de chaque zone.

Une amélioration du modèle hydrique (Lucas, & al., 2003) a été apportée au travers de l'intégration d'un tampon hygroscopique, selon le modèle de Duforestel. Si $n$ est le nombre de zones, un système linéaire de dimension $2n$ est établi et résolu à l'aide d'un schéma aux différences finies.

### 4. ELEMENTS DE VALIDATION DU CODE

La validation du code a été menée en plusieurs étapes dont seules quelques unes sont rappelées ci-dessous. Certaines comparaisons inter logiciels ponctuelles ne sont que mentionnées (CODYBA et TRNSYS pour la partie thermique, BREEZE, COMIS, CONTAM93 pour l'aéraulique sur un cas de l'AIVC TN51), de même que des confrontations analytiques (en aéraulique).

### 4.1. PROCEDURE IEA – BESTEST MONOZONE

CODYRUN a passé la procédure IEA BESTEST Task 12 (cas monozones, séries 600 et 900). La procédure prévoit plus d'une centaine de simulations (Soubdhan & al., 1999) et sur les cas traités, *CODYRUN* affichait des résultats compatibles avec la majorité des programmes de référence, excepté pour quelques uns ayant permis de corriger des erreurs. Un exemple de résultat issu de cette confrontation est le suivant :

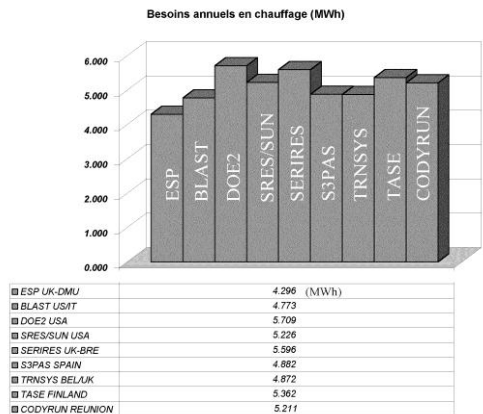

*Figure 1 : Exemple de résultat*

### 4.2. PROCEDURE IEA – BESTEST MULTIZONE

Les cas multizones IEA: SHC Task 34 / ECBCS Annex 43 (Multi-Zone Conduction Cases) (MZ200 -> MZ310) ont été réalisés sans difficulté. Nous avons initié les cas suivants, i.e. les cas MZ320 –> MZ360. Pour le dernier, MZ360 (internal window) il a été nécessaire de prendre en compte les vitrages interzones (et un algorithme de répartition des flux entre zones).

### 4.3. CONFRONTATIONS EXPERIMENTALES

Dans le cadre de campagnes de mesures sur des supports de laboratoire (cellule LGI, STA-TRON, ISOTEST, ETNA EDF-DER) et de logements réels (Trinité, Découverte, ...), un ensemble d'éléments du modèles ont pu être confrontés à des mesures (et pour certains améliorés), principalement vis à vis d'aspects thermiques et hydriques. De manière non exhaustive, certains de ces aspects sont présentés dans les articles (Mara & al., 2001), (Lucas & al., 2001) et (Lauret & al., 2001).

## 5. PRESENTATION DE L'INTERFACE DE L'APPLICATION

Pour des rasions de portabilité, l'application est décomposée en 2 parties : le noyau de calcul est développé en C (pour des raisons de portabilité) alors que l'interface est sous C/Windows. Ce découpage a permis le portage du noyau sous l'environnement TRNSYS (Bastide & al., 2001), sous la forme d'un type autonome.

L'interface du code est assez simple, CODYRUN manipulant des fichiers Bâtiment (extension BTM), des fichiers météorologiques (.MTO, du même format que CODYBA) pour générer des fichiers résultats (dont le contenu est modulable) en mode texte.

En particulier pour la partie de description du bâtiment, cette interface constitue la partie la plus dépouillée de l'application. Il s'agit de fenêtres Windows en mode texte, au sein desquelles toute l'information descriptive (et de choix de modèles) doit être effectuée. Bien que robuste parce qu'éprouvée par de nombreux utilisateurs (étudiants, professionnels et membres de l'équipe), c'est le temps de description qui peut devenir problématique. Pour fixer les idées, le ratio du nombre de zones du bâtiment par rapport aux heures à consacrer à la description (bien que variable selon l'utilisateur) avoisine l'unité.

La structure de fenêtres est l'image de la structure de données de CODYRUN, bien représentée par l'arborescence suivante :

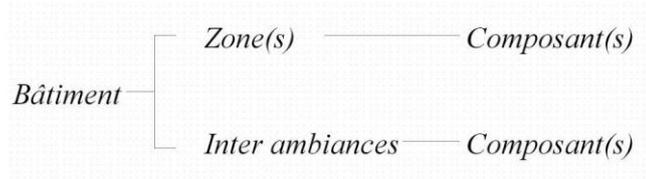

*Figure 2 : Arborescence de données*

Les inter ambiances se définissent comme le lieu d'appartenance des composants séparant 2 zones. Les composants appartiennent à la liste se suivante et se distinguent par leur élément d'appartenance (zone ou inter ambiance) :

| Composants de zone | Composants d'inter ambiance |
|---|---|
| Paroi interne | Paroi de séparation |
| Système de traitement d'air idéal | Terre plein |
| Charge interne | Paroi sur vide sanitaire |
| Bouche de ventilation (VMC) | Paroi sur terre plein |
| | Vitrage |
| | Grande ouverture |
| | Petite ouverture |
| | Débit connu |

*Tableau 1 : Types de composants*

Pour chacun de ces composants des informations spécifiques sont à renseigner pour construire le bâtiment. Un exemple de fenêtre est celle d'un composant de type paroi :

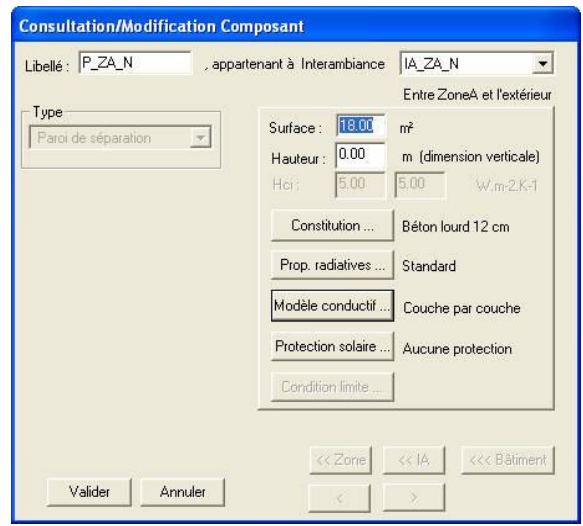

*Figure 3 : Fenêtre principale du composant paroi*

## 6. CONCLUSION

Conduits de façon très autonome et soutenue depuis plus de 10 ans, les développements menés autour de cette application ont conduit notre équipe à disposer d'un environnement de simulation propriétaire et à bâtir en périphérie des thématiques structurantes telles que la validation ou l'application à grande échelle du code dans le cadre de prescriptions architecturales (Garde & al., 2001). Cet environnement évolutif nous a permis de conduire des développements dans des domaines méthodologiques (réduction modale, analyse de sensibilité, couplage avec des algorithmes génétiques, réseaux de neurones, méta modèles, ...) ou liés à des aspects technologiques (masques, intégration de split-system, prise en compte des produits minces réfléchissants, ...).

Accompagné par la croissance de la puissance des machines et faisant l'objet d'un forte demande environnementale (HQE, ...), notre domaine se complexifie de part l'intégration d'autres aspects que ceux initialement liés à la thermique et à l'aéraulique (qualité des ambiances, incluant éclairagisme, polluants, acoustique, ...). Simultanément, il est indispensable de rechercher une meilleure efficacité dans le transfert de connaissances et le caractère opérationnel (i.e. en production) des outils.

Dans sa version actuelle, notre contribution à la communauté est téléchargeable sur www.univ-reunion.fr\iut_dpt_gc, rubrique *Téléchargement*.

## 7. BIBLIOGRAPHIE